\documentclass{ws-procs9x6}

\usepackage{graphicx} 

\begin{document}

\title{COVARIANCE, DYNAMICS AND SYMMETRIES,\\ AND HADRON FORM FACTORS}

\author{M.\ S.\ BHAGWAT, I.\ C.\ CLO\"ET and C.\ D.\ ROBERTS}

\address{Physics Division\\ Argonne National Laboratory\\
             Argonne, IL 60439-4843, U.S.A.}

\begin{abstract}
We summarise applications of Dyson-Schwinger equations to the theory and phenomenology of hadrons.  Some exact results for pseudoscalar mesons are highlighted with details relating to the $U_A(1)$ problem.  We describe inferences from the gap equation relating to the radius of convergence for expansions of observables in the current-quark mass.   We recapitulate upon studies of nucleon electromagnetic form factors, providing a comparison of the $\ln$-weighted ratios of Pauli and Dirac form factors for the neutron and proton.
\end{abstract}

\keywords{Dyson-Schwinger equations; Hadron electromagnetic form factors}

\bodymatter

\section{Introduction}\label{aba:sec1}
\begin{figure}[t]

\centerline{
\includegraphics[clip,width=0.6\textwidth]{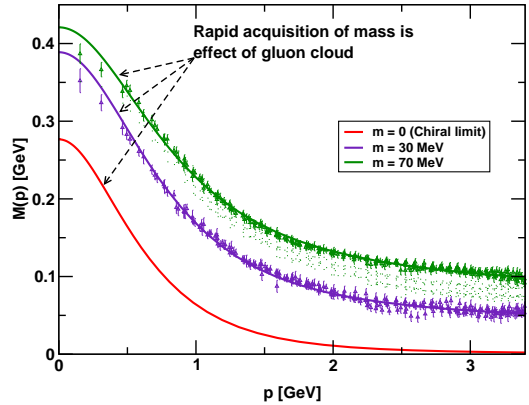}}

\caption{\label{gluoncloud} Dressed-quark mass function, $M(p)$: solid curves -- DSE results, obtained as explained in Refs.\,\protect\cite{Bhagwat:2003vw,Bhagwat:2006tu}, ``data'' -- numerical simulations of unquenched lattice-QCD \protect\cite{Bowman:2005vx}.  In this figure one observes the current-quark of perturbative QCD evolving into a constituent-quark as its momentum becomes smaller.  The constituent-quark mass arises from a cloud of low-momentum gluons attaching themselves to the current-quark.  This is dynamical chiral symmetry breaking: an essentially nonperturbative effect that generates a quark mass \emph{from nothing}; viz., it occurs even in the chiral limit.}
\end{figure}

The Dyson-Schwinger equations (DSEs) are a nonperturbative means of studying QCD in the continuum.  This is illustrated in the overview provided by Ref.\,\cite{Roberts:2007jh}.  A strength of the approach is that the chiral limit; viz., the domain of physical pion masses, is directly accessible.  Indeed, with the established existence of a systematic, nonperturbative and symmetry-preserving truncation scheme \cite{Munczek:1994zz,Bender:1996bb} the DSEs were used to prove that the pion is simultaneously a Goldstone mode and a bound-state of effectively massive constituents \cite{Maris:1997hd}.  Moreover, the exact nature of those dressed-quark and dressed-antiquark constituents is explained by the DSEs \cite{Bhagwat:2003vw,Bhagwat:2006tu}.  Namely, the quark-parton of QCD acquires a momentum-dependent mass function that at infrared momenta is larger by two orders-of-magnitude than the current-quark mass, an effect which owes primarily to a heavy cloud of gluons that clothes a low-momentum quark.\footnote{The $2$-point Schwinger functions for ghosts and gluons experience analogous nonperturbative modification in the infrared \protect\cite{Smekal:1997is}.}  This is illustrated in Fig.\,\ref{gluoncloud}.  

Since a weak coupling expansion of the DSEs generates every diagram in perturbation theory, it is straightforward to ensure that model-dependent assumptions are restricted to infrared momenta.  One can therefore turn a comparison of DSE predictions with data into a probe of the long-range interaction between light-quarks in QCD; namely, a unique means of exploring light-quark confinement.  In large part this study is the same as drawing a map of the infrared behaviour of the QCD $\beta$-function.  It is a fact too often ignored that the potential between infinitely heavy quarks measured in numerical simulations of quenched-QCD -- the static potential -- is not related in any known way to the confinement of light-quarks.  

\section{Gap equation}
Since the gap equation is so fundamental to understanding hadron physics we reproduce it here:
\begin{equation}
S(p)^{-1} =  Z_2 \,(i\gamma\cdot p + m^{\rm bm}) + Z_1 \int^\Lambda_q\! g^2 D_{\mu\nu}(p-q) \frac{\lambda^a}{2}\gamma_\mu S(q) \Gamma^a_\nu(q,p) , \label{gendse}
\end{equation}
where $\int^\Lambda_q$ represents a Poincar\'e invariant regularisation of the integral, with $\Lambda$ the regularisation mass-scale \cite{Maris:1997hd}, $D_{\mu\nu}$ is the dressed-gluon propagator, $\Gamma_\nu$ is the dressed-quark-gluon vertex, and $m^{\rm bm}$ is the quark's $\Lambda$-dependent bare current-mass.  The vertex and quark wave-function renormalisation constants, $Z_{1,2}(\zeta^2,\Lambda^2)$, depend on the gauge parameter.  

The solution; namely, the dressed-quark propagator, can be written 
\begin{eqnarray} 
 S(p) & =&  \frac{1}{i \gamma\cdot p \, A(p^2,\zeta^2) + B(p^2,\zeta^2)} =
\frac{Z(p^2,\zeta^2)}{i\gamma\cdot p + M(p^2)}
%
\label{Sgeneral}
\end{eqnarray} 
wherein the mass function, $M(p^2)$, illustrated in Fig.\,\ref{gluoncloud}, is independent of the renormalisation point, $\zeta$.  The solution is obtained from Eq.\,(\ref{gendse}) augmented by the renormalisation condition
$\left.S(p)^{-1}\right|_{p^2=\zeta^2} = i\gamma\cdot p +
m(\zeta^2)\,,$
where $m(\zeta^2)$ is the running mass: 
$Z_2(\zeta^2,\Lambda^2) \, m^{\rm bm}(\Lambda) = Z_4(\zeta^2,\Lambda^2) \, m(\zeta^2)\,,$
with $Z_4$ the Lagrangian-mass renormalisation constant.  In QCD the chiral limit is strictly defined by \cite{Maris:1997hd}:
$Z_2(\zeta^2,\Lambda^2) \, m^{\rm bm}(\Lambda) \equiv 0 \,, \forall \Lambda^2 \gg \zeta^2 ,$
which states that the renormalisation-point-invariant cur\-rent-quark mass $\hat m = 0$.  In this limit it is possible to unambiguously define the gauge invariant vacuum quark condensate in terms of $S(p)$ \cite{Maris:1997hd,Langfeld:2003ye,Chang:2006bm}.  This fact emphasises that gauge covariant quantities contain gauge invariant information. 

The question of whether $M(p^2)$ has an expansion in current-quark mass around its chiral-limit value bears upon the radius of convergence for chiral perturbation theory.  With this question one is asking whether it is possible to write
\begin{equation}
M(p^2;\hat m) = M(p^2;\hat m=0) + \sum_{n=1}^\infty a_n \hat m^n
\end{equation}
on a measurable domain of current-quark mass.  It was found \cite{Chang:2006bm} that such an expansion exists for $\hat m < \hat m_{\rm rc}$.  The value of $\hat m_{\rm rc}$ can be reported as follows: for a pseudoscalar meson constituted from a quark, $f$, with mass $\hat m_{\rm rc}$ and its antiparticle, $\bar f$, $m_{\bar f f}^{0^-} = 0.45\,$GeV.  Since physical observables, such as the leptonic decay constant of a pseudoscalar meson, are expressed in terms of $M(p^2)$, it follows that a chiral expansion is meaningful only for $(m_{\bar f f}^{0^-})^2 \lesssim 0.2\,$GeV$^2$.\footnote{NB.\ Irrespective of the current-mass of the other constituent, a pseudoscalar meson containing one current-quark whose mass exceeds $\hat m_{\rm cr}$ is never within the domain of uniform convergence.}  This entails, e.g., that it is only valid to employ chiral perturbation theory to fit and extrapolate results from numerical simulations of lattice-regularised QCD when the simulation parameters provide for $m_\pi^2 \lesssim 0.2\,$GeV$^2$.  Lattice results at larger pion masses are not within the domain of convergence of chiral perturbation theory. 

\section{Mesons and the axial-vector Ward-Takahashi identity}
Dynamical chiral symmetry breaking is a fact in QCD and this amplifies the importance of the axial-vector Ward-Takahashi identity.  The existence of a sensible truncation of the DSEs has enabled proof via that identity of a body of exact results for pseudoscalar mesons.  They relate even to radial excitations and/or hybrids \cite{Holl:2004fr,Holl:2005vu,McNeile:2006qy}, and heavy-light \cite{Ivanov:1998ms} and heavy-heavy mesons \cite{Bhagwat:2006xi}.  The results have been illustrated using a renormalisation-group-improved rainbow-ladder truncation \cite{Maris:1997tm,Maris:1999nt}, which also provided a prediction of the electromagnetic pion form factor \cite{Maris:2000sk}.

Implications for neutral pseudoscalar mesons have been elucidated \cite{Bhagwat:2007ha}.  In the general case the axial-vector Ward-Takahashi identity is written 
\begin{equation}
P_\mu \Gamma_{5\mu}^a(k;P) ={\cal S}^{-1}(k_+) i \gamma_5 {\cal F}^a 
+ i \gamma_5 {\cal F}^a {\cal S}^{-1}(k_-)
- 2 i {\cal M}^{ab}\Gamma_5^b(k;P)  - {\cal A}^a(k;P)\,,
\label{avwti}
\end{equation}
wherein: $\{{\cal F}^a | \, a=0,\ldots,N_f^2-1\}$ are the generators of $U(N_f)$; the dressed-quark propagator ${\cal S}=\,$diag$[S_u,S_d,S_s,S_c,S_b,\ldots]$ is matrix-valued; ${\cal M}(\zeta)$ is the matrix of renormalised (running) current-quark masses and
${\cal M}^{ab} = {\rm tr}_F \left[ \{ {\cal F}^a , {\cal M} \} {\cal F}^b \right],$
where the trace is over flavour indices.  The inhomogeneous axial-vector vertex in Eq.\,(\ref{avwti}), $\Gamma_{5\mu}^a(k;P)$, where $P$ is the total momentum of the quark-antiquark pair and $k$ the relative momentum, satisfies a Bethe-Salpeter equation, and likewise the pseudoscalar vertex, $\Gamma_5^b(k;P)$.

The final term in Eq.\,(\ref{avwti}) expresses the non-Abelian axial anomaly.  It involves 
\begin{equation}
{\cal A}_U(k;P) = \!\!  \int\!\! d^4xd^4y\, e^{i(k_+\cdot x - k_- \cdot y)} N_f \left\langle  {\cal F}^0\,q(x)  \, {\cal Q}(0) \,   \bar q(y) 
\right\rangle, \label{AU}
\end{equation}
wherein the matrix element represents an operator expectation value in full QCD and
${\cal Q}(x) = i \frac{\alpha_s }{8 \pi} \, \epsilon_{\mu\nu\rho\sigma} F^a_{\mu\nu} F^a_{\rho\sigma}(x) 
= \partial_\mu K_\mu(x)$
is the topological charge density operator, where $F_{\mu\nu}^a$ is the gluon field strength tensor.  It is plain and important that only ${\cal A}^{a=0}$ is nonzero. 

It is fundamentally important that while ${\cal Q}(x)$ is gauge invariant, the associated Chern-Simons current, $K_\mu$, is not.  It follows that in QCD no physical state can couple to $K_\mu$ and hence that physical states cannot provide a resolution of the so-called $U_A(1)$-problem; namely, they cannot play any role in ensuring that the $\eta^\prime$ is not a Goldstone mode.

In thinking about the $U_A(1)$-problem it is only necessary to focus on the case ${\cal A}^{a=0} \neq 0$ because if that is not so, then following Ref.\,\cite{Maris:1997hd} it is clear that the $\eta^\prime$ is certainly a Goldstone mode.  ${\cal A}^0(k;P)$ is a pseudoscalar vertex and can therefore be expressed
\begin{eqnarray}
\nonumber 
{\cal A}^0(k;P) & = & {\cal F}^0\gamma_5 \left[ i {\cal E}_{\cal A}(k;P) + \gamma\cdot P {\cal F}_{\cal A}(k;P)  \right. \\
&& \left. +\gamma\cdot k k\cdot P {\cal G}_{\cal A}(k;P) + \sigma_{\mu\nu} k_\mu P_\nu {\cal H}_{\cal A}(k;P)\right].
\end{eqnarray}
Equation\,(\ref{avwti}) can now be used to derive a collection of chiral-limit, pointwise Goldberger-Treiman relations, important amongst which is the identity
\begin{equation}
\label{ewti}
2 f_{\eta^\prime} E_{\eta^\prime}(k;0) = 2 B_{0}(k^2) - {\cal E}_{\cal A}(k;0)\,,
\end{equation}
where $B_0(k^2)$ is obtained in solving the chiral-limit gap equation.

It is now plain that if 
\begin{equation}
\label{calEB}
{\cal E}_{\cal A}(k;0) = 2 B_{0}(k^2) \,,
\end{equation}
then $f_{\eta^\prime} E_{\eta^\prime}(k;0) \equiv 0$.  This being true, then the homogeneous Bethe-Salpeter equation for the $\eta^\prime$ does not possess a massless solution in the chiral limit.  The converse is also true; namely, the absence of such a solution requires Eq.\,(\ref{calEB}).  Hence,  Eq.\,(\ref{calEB}) is a necessary and sufficient condition for the absence of a massless $\eta^\prime$ bound-state.  It is the chiral limit that is being discussed, in which case $B_{0}(k^2) \neq 0$ if, and only if, chiral symmetry is dynamically broken.   Hence, the absence of a massless $\eta^\prime$ bound-state is only assured through the existence of an intimate connection between DCSB and an expectation value of the topological charge density.  

Reference\,\cite{Bhagwat:2007ha} derives a range of corollaries, amongst which are mass formulae for neutral pseudoscalar mesons, and presents an \textit{Ansatz} for the Bethe-Salpeter kernel that enables their illustration.  Despite its simplicity, the model is elucidative and phenomenologically efficacious; e.g., it predicts $\eta$--$\eta^\prime$ mixing angles of $\sim - 15^\circ$, $\pi^0$--$\eta$ angles of $\sim 1^\circ$, and a strong neutron-proton mass difference of $0.75\,(\hat m_d - \hat m_u)$.

\section{Baryons}
Partly because attention to this sector has only recently increased, the current level of expertise in studying baryons is roughly the same as it was with mesons more than ten years ago; viz., model building and phenomenology.  We are a little ahead of that game, however, because much has been learnt in meson applications; e.g., as outlined above, a veracious understanding of the structure of dressed-quarks and -gluons has been acquired.   

In quantum field theory a nucleon appears as a pole in a six-point quark Green function.  The pole's residue is proportional to the nucleon's Faddeev amplitude, which is obtained from a Poincar\'e covariant Faddeev equation that adds-up all possible quantum field theoretical exchanges and interactions that can take place between three dressed-quarks.  This is important because modern, high-luminosity experimental facilities employ large momentum transfer reactions; viz., $Q^2 > M_N^2$ where $M_N$ is the nucleon's mass, and hence a veracious understanding of contemporary data requires a Poincar\'e covariant description of the nucleon.  

A tractable truncation of the Faddeev equation is based \cite{Cahill:1988dx} on the observation that an interaction which describes mesons also generates diquark correlations in the colour-$\bar 3$ channel \cite{Cahill:1987qr}.  The dominant correlations for ground state octet and decuplet baryons are $0^+$ and $1^+$ diquarks because, e.g.: the associated mass-scales are smaller than the baryons' masses \cite{Burden:1996nh,Maris:2002yu}, namely (in GeV) -- 
$m_{[ud]_{0^+}} = 0.74 - 0.82$,
$m_{(uu)_{1^+}}=m_{(ud)_{1^+}}=m_{(dd)_{1^+}}=0.95 - 1.02$;
and the electromagnetic size of these correlations is less than that of the proton \cite{Maris:2004bp} -- $r_{[ud]_{0^+}} \approx 0.7\,{\rm fm}$, which implies  $r_{(ud)_{1^+}} \sim 0.8\,{\rm fm}$ based on the $\rho$-meson/$\pi$-meson radius-ratio \cite{Maris:2000sk,Bhagwat:2006pu}.

The Faddeev equation's kernel is completed by specifying that the quarks are dressed, with two of the three dressed-quarks correlated always as a colour-$\bar 3$ diquark.  Binding is then effected by the iterated exchange of roles between the bystander and diquark-participant quarks.  A Ward-Takahashi-identity-pre\-ser\-ving electromagnetic current for the baryon thus constituted is subsequently derived~\cite{Oettel:1999gc}.  It depends on the electromagnetic properties of the axial-vector diquark correlation.

A study of the nucleon's mass and the effect on this of a pseudoscalar meson cloud are detailed in \cite{Hecht:2002ej}.  Lessons learnt were employed in a series of studies of nucleon properties, including form factors \cite{Alkofer:2004yf,Holl:2005zi,Bhagwat:2006py,Holl:2006zw}.  The calculated ratio $\mu_p G_E^p(Q^2)/G_M^p(Q^2)$ passes through zero at $Q^2\approx 6.5\,$GeV$^2$ \cite{Holl:2005zi}.  The analogous ratio for the neutron was calculated \cite{Bhagwat:2006py}.  In the neighbourhood of $Q^2=0$, $\mu_n\, G_E^n(Q^2)/G_M^n(Q^2) = - \frac{r_n^2}{6}\, Q^2$, where $r_n$ is the neutron's electric radius.  The evolution of $\mu_p G_E^p(Q^2)/G_M^p(Q^2)$ and $\mu_n G_E^n(Q^2)/G_M^n(Q^2)$ on $Q^2\gtrsim 2\,$GeV$^2$ are both primarily determined by the quark-core of the nucleon.  While the proton ratio decreases uniformly on this domain \cite{Alkofer:2004yf,Holl:2005zi},  the neutron ratio increases steadily until $Q^2\simeq 8\,$GeV$^2$ \cite{Bhagwat:2006py}.  

\begin{figure}[t]

\centerline{
\includegraphics[clip,width=0.6\textwidth]{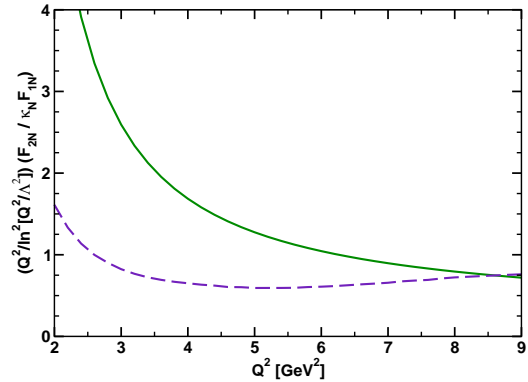}}

\caption{\label{nF2F1} Weighted nucleon Pauli/Dirac form factor ratio, calculated using the framework of Ref.\,\protect\cite{Alkofer:2004yf} and presented with $\Lambda=0.94$GeV: solid curve -- neutron; dashed curve -- proton.
}
\end{figure}

A form factor ratio motivated by ideas from perturbative QCD \cite{Belitsky:2002kj} is depicted in Fig.\,\ref{nF2F1}.  The parameter $\Lambda$ is interpreted as a mass-scale that defines the upper-bound on the domain of so-called soft momenta in the perturbative analysis.  A plausible value for such a quantity is $\Lambda \sim M_N$ \cite{Alkofer:2004yf}.\footnote{NB.\ A value of $\Lambda \sim 0.3\,$GeV corresponds to a length-scale $r_\Lambda \sim 1\,$fm and it is not credible that perturbative QCD is applicable at ranges greater than the proton's radius.}  It is interesting that the model of Ref.\,\cite{Alkofer:2004yf} yields neutron and proton ratios which cross at $Q^2\simeq 8.5\,$GeV$^2$.  The model's prediction for truly asymptotic momenta is currently being explored.

Stimulated by the possibility that Nature's fundamental ``constants'' might actually exhibit spatial and temporal variation \cite{Uzan:2002vq}, we have begun to explore the current-quark-mass-dependence of the nucleon magnetic moments.  This complements work on hadron $\sigma$-terms \cite{Flambaum:2005kc,Holl:2005st,Flambaum:2007mj}.  Preliminary results for the quark-core contribution to this variation can be expressed through the following ratios evaluated at the physical current-quark mass:
\begin{equation}
\begin{array}{c|c|c}
N & p & n \\\hline
\rule{0em}{2.5ex}-\frac{\delta \mu_N}{\mu_N} / \frac{\delta m}{m} &  ~0.016~ & ~0.0042~
\end{array}
\end{equation}
It is likely that pseudoscalar meson contributions will increase these values by a factor of $\gtrsim 10$ \cite{Flambaum:2004tm}.

\section{Epilogue}
Confinement and dynamical chiral symmetry breaking (DCSB) can only be veraciously understood in relativistic quantum field theory.  The DSEs provide a natural vehicle for the exploration of these phenomena.  

DCSB is a singularly effective mass generating mechanism.  For light-quarks it far outweighs the Higgs mechanism.  It is understood via QCD's gap equation, which delivers a quark mass function with a momentum-dependence that connects the perturbative domain with the nonperturbative, con\-sti\-tuent-quark domain.  

The existence of a sensible truncation scheme enables the proof of exact results using the DSEs.  That the truncation scheme is also tractable means the results can be illustrated, and furnishes a practical tool for the prediction of observables.  The consequent opportunities for rapid feedback between experiment and theory brings within reach an intuitive understanding of nonperturbative strong interaction phenomena.  

It can be argued that confinement is expressed in the analyticity properties of dressed Schwinger functions.  To build understanding it is essential to work toward an accurate map of the confinement force between light-quarks and elucidate if/how that evolves from the potential between two static quarks.  Among the rewards are a clear connection between confinement and DCSB, an accounting of the distribution of mass within hadrons, and a realistic picture of hybrids and exotics.

It is important to understand the relationship between parton properties on the light-front and the rest frame structure of hadrons.  This is a challenge because, e.g., it is difficult to see how DCSB, a keystone of low-energy QCD, can be realised on the light-front.  Parton distribution functions must be calculated in order to learn their content.  Parametrisation is insufficient.  It would be very interesting to know how, if at all, the distribution functions of a Goldstone mode differ from those of other hadrons.  Answers to these and kindred questions are being sought using the DSEs \cite{Hecht:2000xa,Cloet:2007em}.

\section*{Acknowledgments}
We acknowledge useful communications with 
V.\,V.~Flambaum and A.~H\"oll.
This work was supported by the Department of Energy, Office of Nuclear Physics, contract no.\ DE-AC02-06CH11357; and benefited from the facilities of the ANL Computing Resource Center.

\end{document}